\documentclass{PoS}
\pdfoutput=1

\title{LHC Prospects for Asymmetric Top-Antitop Production}

\ShortTitle{LHC Prospects for Asymmetric Top-Antitop Production}

\author{Susanne Westhoff\\
        PITTsburgh Particle physics, Astrophysics \& Cosmology Center (PITT-PACC)\\ 
Department of Physics \& Astronomy\\
        University of Pittsburgh\\
        Pittsburgh, PA 15260, USA\\
        E-mail: \email{suw22@pitt.edu}}

\abstract{The charge asymmetry in hadronic top-antitop-quark production is a powerful test observable of perturbative quantum chromodynamics. Since its origin in terms of Feynman diagrams is distinct from charge-symmetric top-antitop production, probing the charge asymmetry provides complementary information to cross section measurements. Furthermore, asymmetric observables offer new ways to distinguish new physics beyond the standard model. This article summarizes prospects to investigate the top-quark charge asymmetry at the LHC in and beyond the standard model. Particular attention is given to observables in top-antitop production in association with a jet, a photon or a $W$ boson.}

\FullConference{Frontiers of Fundamental Physics 14 - FFP14,\\
                15-18 July 2014\\
                Aix Marseille University (AMU) Saint-Charles Campus, Marseille\hfill {\rm PITT-PACC-1501}}
		
\begin{document}

\section{Introduction and state of the art}
The charge asymmetry in hadronic top-antitop production has recently attracted much attention for two good reasons. Firstly, in quantum chromodynamics (QCD) the asymmetry is absent at the leading order (LO), which makes it not only an important test of perturbative quantum field theory, but also a sensitive probe of physics beyond the standard model (SM). Secondly, the experi\-mental progress at high-energy hadron colliders, namely the Tevatron and the LHC, allows us to compare predictions of observables in top-pair production with reality to good precision.

Generally, probing the charge asymmetry in top-antitop production means comparing a specific kinematic constellation of tops and antitops with the same constellation where tops and antitops are interchanged. At the parton level, the charge asymmetry can be defined in terms of the scattering angle $\theta$ of the heavy quark with momentum $p_1$ with respect to the incoming light quark inside the proton in the partonic center-of-mass (CM) frame,
\begin{equation}\label{eq:charge-asymmetry}
\frac{d\sigma_A}{d\cos\theta} = \frac{d\sigma_{t\bar t}(t(p_1)\,\bar t(p_2))}{d\cos\theta} - \frac{d\sigma_{t \bar t}(\bar t(p_1)\,t(p_2))}{d\cos\theta}.
\end{equation}
Most observables test this angular asymmetry directly or in parts through rapidity differences. At the Tevatron, the charge asymmetry has been measured directly as a top-quark forward-backward asymmetry $A_{\rm FB} = \int_0^1 d\cos\theta\, d\sigma_A/\sigma_{t\bar t}$, where $\sigma_{t\bar t}$ is the total cross section of inclusive top-antitop production~\mbox{\cite{Aaltonen:2012it,Abazov:2014cca}}. An asymmetry measurement at the LHC in terms of rapidity differences gives only partial access to the angular charge asymmetry in (\ref{eq:charge-asymmetry}) (see Section~\ref{sec:lhc} and \cite{Kuhn:2011ri} for an overview).  Alternatively the charge asymmetry can be measured at the LHC in terms of the energy difference between top and antitop quarks (see Section~\ref{sec:ea}).

Much of the thorough investigation of the top charge asymmetry has been stimulated by observed discrepancies at the Tevatron experiments. The potential of new physics contributing to the top charge asymmetry has led to a variety of new analyses of the top-quark sector, in particular with LHC observables (for comprehensive reviews see \cite{Kamenik:2011wt,Westhoff:2011tq,Aguilar-Saavedra:2014kpa}). These searches have resulted in strong bounds on scenarios that can give rise to the large enhancement of the asymmetry suggested by the original Tevatron measurements.

\begin{figure}[!t]
\centering
\begin{tabular}{cc}
\hspace*{-0.2cm}\raisebox{0.1cm}{\includegraphics[scale=0.82]{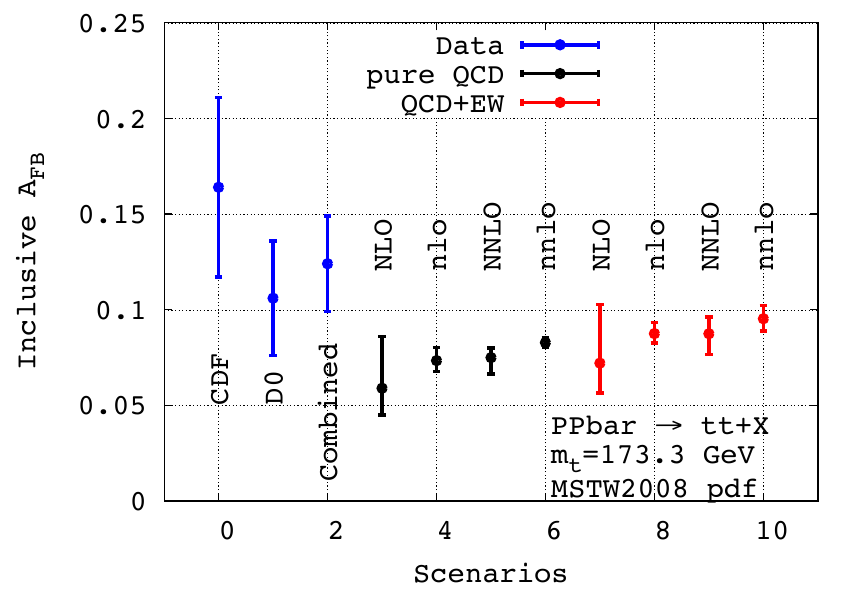}}\hspace*{0.1cm} & \includegraphics[scale=0.87]{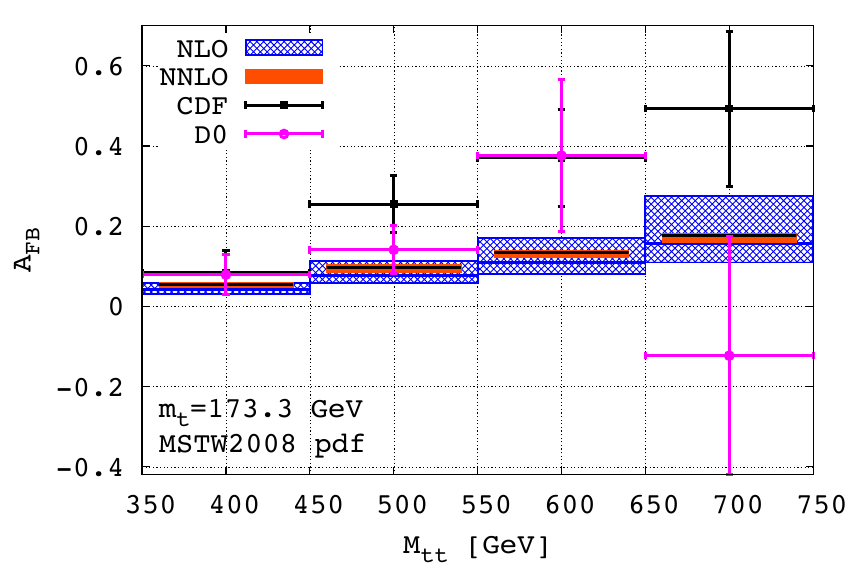}
\end{tabular}
\caption{\label{fig:tevatron}Measurements of the top-quark forward-backward asymmetry at the Tevatron compared to the SM prediction~\cite{Czakon:2014xsa}.
 Left: total forward-backward asymmetry $A_{\rm FB}$. Right: differential asymmetry in the top-antitop invariant mass $m_{t\bar t}$.}
\end{figure}

The originally observed excess in several asymmetry measurements at the Tevatron has been largely resolved by now. On the theory side, electroweak contributions~\cite{Hollik:2011ps} and next-to-next-to-leading-order (NNLO) corrections in QCD have been shown to enhance the leading asymmetry in QCD to $A_{\rm FB}^{\rm SM} = (9.5\pm 0.7)\%$~\cite{Czakon:2014xsa}. On the experimental side, analyses of the complete Tevatron data set have resulted in a better agreement with the SM prediction. Figure~\ref{fig:tevatron} summarizes the latest Tevatron measurements of the asymmetry and compares them to up-to-date SM predictions. The left panel shows the total forward-backward asymmetry $A_{\rm FB}$. A modest excess of the CDF measurement persists, but the combination of CDF and D0 measurements agrees with the SM prediction within one standard deviation. The right panel displays the differential asymmetry as a function of the top-antitop invariant mass $m_{t\bar t}$. It is apparent that the difference between the CDF and D0 measurements of $A_{\rm FB}$ is to a good extent due to the discrepancy in the highest $m_{t\bar t}$ range. In the region of lower $m_{t\bar t}$, both measurements agree within one standard deviation and show only a small enhancement over the SM prediction.

With the LHC warmed up and soon running at full power, the charge asymmetry reaches a new sensitivity level for testing QCD and probing new physics. The remainder of this review focuses on opportunities to test the SM and new physics through charge asymmetry observables in top-antitop production at the LHC. In Section~\ref{sec:lhc}, we summarize the rapidity asymmetry measurements during the first run of the LHC and discuss the observation prospects for run 2. In Section~\ref{sec:ea}, an alternative observable called energy asymmetry is suggested, which enhances the sensitivity to the charge asymmetry in top-antitop production and helps to overcome experimental limitations from charge-symmetric background. In Sections~\ref{sec:jet-photon} and \ref{sec:ttw} we finally argue how the radiation of an additional jet, photon or $W$ boson in top-antitop production can be used in the search for new physics.

\section{Rapidity asymmetry at the LHC}\label{sec:lhc}
\begin{figure}[!t]
\centering
\begin{tabular}{cc}
\includegraphics[scale=0.32]{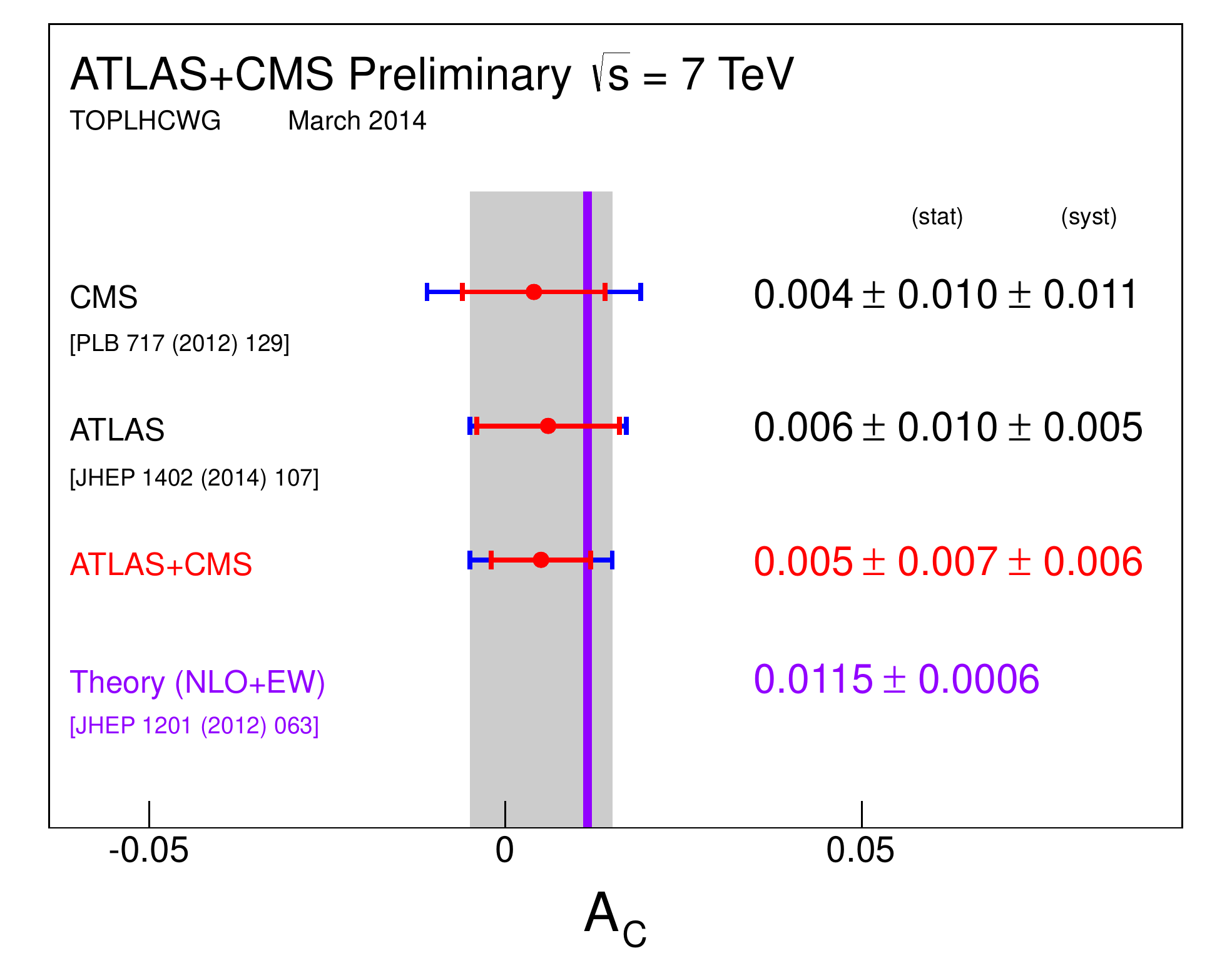}\hspace*{0.1cm} & \raisebox{0.2cm}{\includegraphics[scale=0.9]{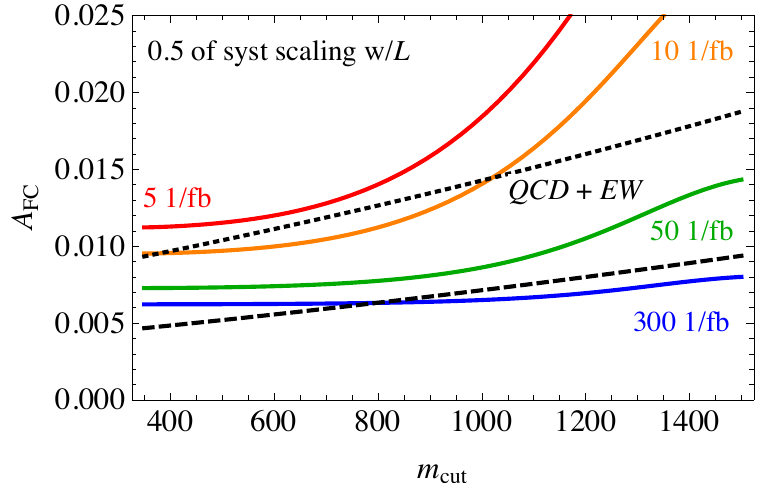}}
\end{tabular}
\caption{\label{fig:lhc}Measurements of the inclusive top-quark charge asymmetry at the LHC compared to SM prediction. Left: combined ATLAS and CMS results for the inclusive asymmetry $A_C$~\cite{LHC-combi}. Right: Projected uncertainties (in color) on a LHC measurement of $A_C$ at $\sqrt{s}=14\,{\rm TeV}$ with a lower invariant mass cut $m_{\rm cut}=m_{t\bar t}^{\rm min}$~\cite{Jung:2013vpa}. The SM prediction $A_C^{\rm SM}$ ($0.5\times A_C^{\rm SM}$) is displayed as a dotted (dashed) black curve.}
\end{figure}

At the LHC the measurement of the top charge asymmetry is challenged by large background from the partonic process $gg\to t\bar t +X$. During run 1, the ATLAS and CMS collaborations have investigated the asymmetry in terms of top-antitop rapidity differences,
\begin{equation}\label{eq:ac}
A_C = \frac{\sigma_{t\bar t}(\Delta |y| > 0)-\sigma_{t\bar t}(\Delta |y| < 0)}{\sigma_{t\bar t}(\Delta |y| > 0)+\sigma_{t\bar t}(\Delta |y| < 0)},
\end{equation}
with the difference of absolute top and antitop rapidities, $\Delta |y| = |y_t|-|y_{\bar t}|$, in the laboratory frame. The results are displayed in the left panel of Figure~\ref{fig:lhc}. While these measurements are in agreement with the SM prediction, they are also consistent with a non-observation of an asymmetry. The discovery of the top charge asymmetry is thus handed over to run 2. The significance of a signal will ultimately depend on the control of systematic errors. In Figure~\ref{fig:lhc}, right, the asymmetry $A_{\rm FC} \equiv A_C$ in proton-proton collisions at a CM energy of $\sqrt{s}=14\,{\rm TeV}$ is shown as a function of a lower cut on the top-antitop invariant mass, $m_{t\bar t}^{\rm min}$. Assuming that at least half of the systematic uncertainties during run 1 scale down with the luminosity, the LHC may ultimately be sensitive to the rapidity asymmetry at the $95\%$~CL.

In order to benefit from the charge asymmetry as a powerful test of QCD and a probe of new physics, the goal is to measure it at the LHC with a better precision than at the Tevatron. The precise theory prediction of the Tevatron asymmetry, $A_{\rm FB}^{\rm SM} = (9.5\pm 0.7)\%$, suggests that LHC asymmetry observables can also be predicted at the percent level. The main challenge is thus to overcome the experimental obstacles by developing new observables that can be measured with percent precision, too.  

One road pursued in this respect is to consider top-antitop production in association with an additional jet, $pp\to t\bar t + j$. In this process the charge asymmetry in QCD is generated at the LO, which results in a larger rapidity asymmetry than in inclusive top-antitop production. Part of the asymmetry at LO, however, is cancelled by large negative corrections at the next-to-leading order (NLO)~\cite{Dittmaier:2007wz,Melnikov:2010iu,Alioli:2011as}, which makes an observation again very difficult. A strong kinematical cut on the region where the jet is emitted perpendicular to the beam line can help to facilitate a measurement with a sufficiently large data set~\cite{Berge:2012rc}.

\section{Energy asymmetry in top-pair production with jet association}\label{sec:ea}
The previously discussed observables probe the charge asymmetry in the quark-antiquark channel $q\bar q\to t\bar t +X$, which however occurs in less than $5\%$ of the produced top-antitop pairs in proton-proton collisions at $\sqrt{s}=14\,{\rm TeV}$. In order to overcome the large gluon-gluon background at the LHC, it is useful to rather focus on the quark-gluon channel $qg\to t\bar t +X$, which induces more than $20\%$ of the produced top-antitop pairs. In inclusive production, the asymmetry from quark-gluon collisions is negligibly small~\cite{Kuhn:1998kw}. However, if an additional quark-jet is observed, as in $qg\to t\bar t q$, this jet handle can be exploited to define asymmetry observables beyond rapidity differences.

\begin{figure}[!t]
\centering
\begin{tabular}{cc}
\hspace*{-0.1cm}\raisebox{1.7cm}{\includegraphics[scale=0.3]{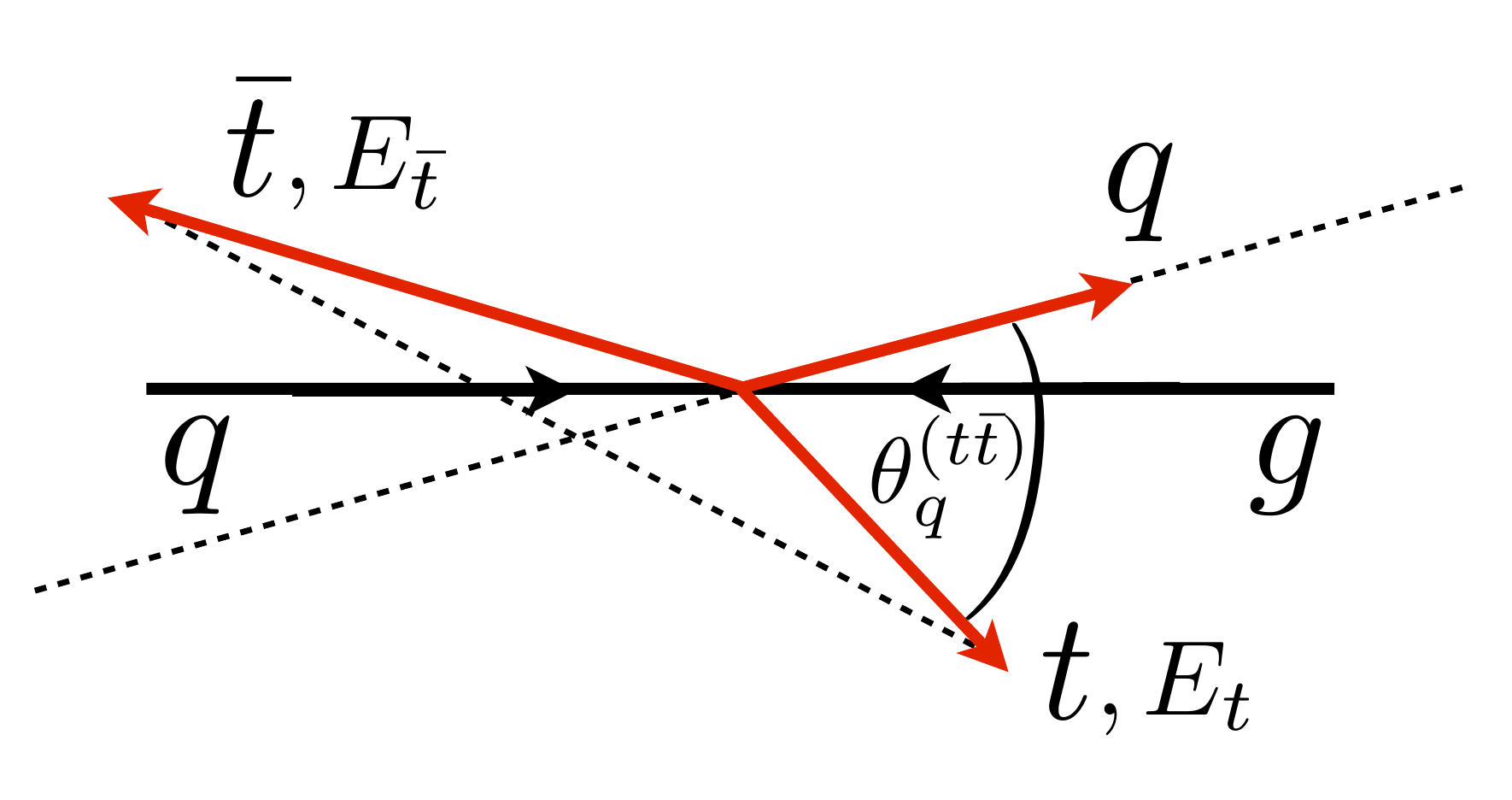}}\hspace*{0.9cm} & \includegraphics[scale=0.7]{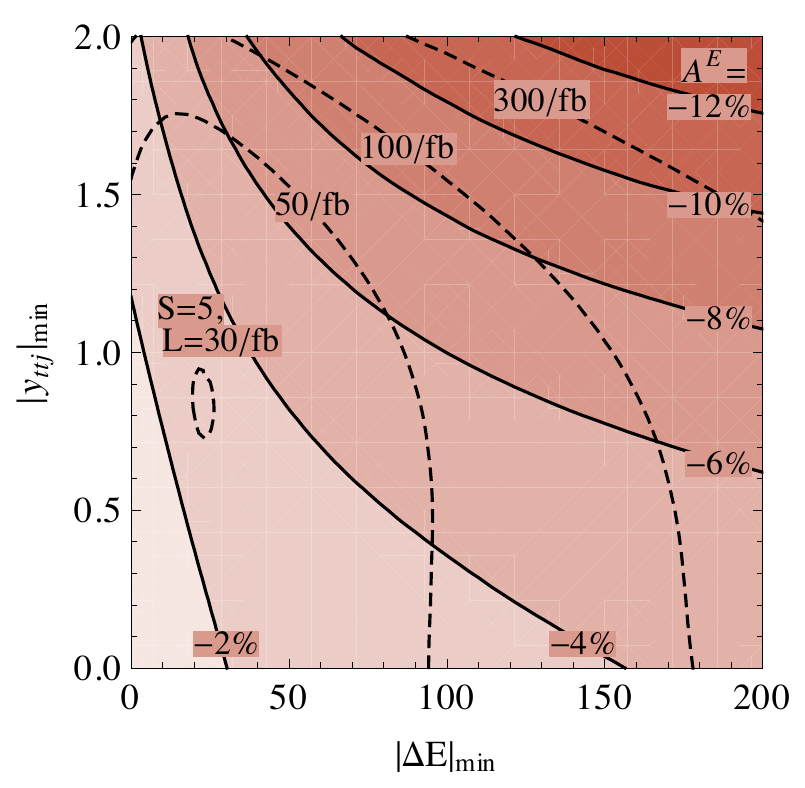}
\end{tabular}
\caption{\label{fig:ea}Energy asymmetry $A_E$ at the LHC with $\sqrt{s}=14\,{\rm TeV}$~\cite{Berge:2013xsa}. Left: illustration of kinematics in quark-gluon parton collisions. Right: $A_E$ (black curves) as a function of lower cuts on the top-antitop energy difference, $|\Delta E|_{\rm min}$, and on the boost of the final state, $|y_{t\bar t j}|_{\rm min}$. A cut on the jet rapidity, $|y_j| < 0.5$, has been applied. The statistical significance of the signal, $S=|A_E|/\delta A_E=5$, is shown for various luminosities $L$ (dashed curves).}
\end{figure}

A particularly useful observable is the energy asymmetry in top-antitop plus jet production~\cite{Berge:2013xsa}, defined as
\begin{equation}\label{eq:ea}
A_{E} = \frac{\sigma_{t\bar t}(\Delta E > 0)-\sigma_{t\bar t}(\Delta E < 0)}{\sigma_{t\bar t}(\Delta E > 0)+\sigma_{t\bar t}(\Delta E < 0)},
\end{equation}
in terms of the difference between the top and antitop energies in each event, $\Delta E = E_t - E_{\bar t}$. The energies are to be taken in the partonic CM frame. The energy asymmetry in the quark-gluon channel corresponds with a forward-backward asymmetry of the quark-jet in the rest frame of the top-antitop pair. This kinematical connection, which relies solely on momentum conservation in the final state, is illustrated in Figure~\ref{fig:ea}, left. The measured quantity is thus similar to the Tevatron asymmetry, namely the forward-backward asymmetry of the top-quark in the quark-antiquark rest frame. In particular, the energy asymmetry gives direct access to the charge asymmetry, unlike the LHC rapidity asymmetry $A_C$ from (\ref{eq:ac}). The definition of $A_E$ in (\ref{eq:ea}) corresponds to the sum of asymmetries in the $qg$ and $gq$ channels, whereas contributions from the $q\bar q$ and $\bar q q$ channels cancel in the sum. 

In the full data set of $t\bar t + j$ production at $\sqrt{s}=14\,{\rm TeV}$ the energy asymmetry is tiny, $A_E=0.5\%$. However, experimental cuts on three kinematical quantities can enhance the asymmetry to an observable level. First, the gluon-gluon background can be efficiently reduced by focusing on boosted $t\bar t + j$ events. Second, the sum of $qg$ and $gq$ contributions reaches its maximum when the jet is emitted perpendicular to the beam axis. And third, the asymmetry grows with the energy difference $|\Delta E|$. In Figure~\ref{fig:ea}, right, we show the effect of these cuts on the energy asymmetry $A_E$. With suitable cuts, the asymmetry can reach up to $A_E\approx -12\%$ (solid black lines). The dashed contours correspond to a statistical significance of five standard deviations for a given event luminosity, assuming an experimental efficiency of $5\%$. The observability of the energy asymmetry will benefit from a larger data set, where stronger cuts can enhance the observable over systematic uncertainties. The prospects to observe the top charge asymmetry at the LHC first through the energy asymmetry in $t\bar t + j$ production are thus very promising. Investigations of the experimental implementation and a refined theory prediction, including NLO corrections, are underway.

\section{New-physics searches with a jet or photon handle}\label{sec:jet-photon}
Exploring new asymmetry observables at the LHC can be a powerful tool to test scenarios beyond the SM. From a theory perspective, the charge asymmetry in top-antitop production is complementary to inclusive top-antitop production in that it probes virtual particles with different couplings to quarks. For instance, the charge asymmetry is most sensitive to axial-vector currents, whereas top-antitop-symmetric observables test dominantly vector currents. While symmetric observables can typically be measured with better precision, charge asymmetries are a valuable tool to pin down the coupling structure of new currents and find new particles that hide in other observables.

In this section, we present LHC opportunities to search for new physics via asymmetry observables in top-antitop production in association with an additional jet or photon. Given the large uncertainty on the current rapidity asymmetry measurements both at the Tevatron and the LHC, novel asymmetry observables have a considerable parameter space yet to explore. Since heavy resonances are generally strongly constrained by LHC searches during run 1, we focus on new particles with masses below the top-antitop production threshold. The example of a massive color-octet vector boson with axial-vector couplings to quarks (commonly dubbed ``axigluon'') will illustrate the sensitivity of asymmetry observables to new physics. The axigluon is known to induce a large forward-backward asymmetry at the Tevatron, while hiding in other observables~\cite{Tavares:2011zg,Gresham:2012kv,Haisch:2011up}.

An additional jet in top-antitop production broadens the possibilities to test the features of a new particle beyond inclusive top-antitop production. In addition to probing axial-vector coup\-lings, the charge asymmetry in $t\bar t + j$ production is sensitive to new vector currents at the tree level~\cite{Ferrario:2009ee}. In particular, the angular distribution of the jet (the ``jet handle'') is a good analyzer of the origin of the asymmetry~\cite{Berge:2012rc}. If the asymmetry is generated by the interference of initial- and final-state radiation, the amplitude is maximized for jet emission perpendicular to the beam line. In turn, if the asymmetry does not originate from the presence of the jet (but for instance from the interference of a vector and an axial-vector current in top-antitop production), the amplitude is enhanced for jet emission collinear to the beam line. With suitable cuts on the jet direction, both the rapidity asymmetry and the energy asymmetry can be optimized to observe specific new coupling structures. It has been shown that for the light axigluon scenario suggested in~\cite{Tavares:2011zg}, the entire parameter space can be tested during LHC run 2 with rapidity and energy asymmetries in $t\bar t + j$ production~\cite{Alte:2014toa}.

The rapidity asymmetry in $t\bar t + \gamma$ production has been invoked as another useful tool to distinguish new physics in top-antitop production. Compared to inclusive top-antitop production, the presence of an additional photon increases the number of quark-antiquark events and thereby enhances the charge asymmetry over the gluon-gluon background. Furthermore, the photon coupling is sensitive to new interactions that violate weak isospin symmetry. In particular, a scenario with different couplings to up- and down-quarks typically leads to very different effects on the rapi\-di\-ty asymmetries in $t\bar t$ and $t\bar t + \gamma$ production~\cite{Aguilar-Saavedra:2014vta}. While a measurement of $A_C$ in $t\bar t + \gamma$ events in data collected during run 1 is strongly limited by statistics, a good level of model discrimination is expected with more data during run 2.

\section{Charge asymmetry in $t\bar t + W$ boson production}\label{sec:ttw}
Top-antitop production with an additional $W$ boson differs from $t\bar t + j$  and $t\bar t + \gamma$ production by involving a charged weak current. Since the charged current in $t\bar t + W$ production must occur in the initial state, the process is purely induced by a quark-antiquark state at the tree level. The rapidity asymmetry is thus larger than in inclusive top-antitop production, which is dominantly gluon-gluon-induced~\cite{Maltoni:2014zpa}. A second important feature is the polarization of the top-quarks due to the initial left-handed weak current. The top polarization is perfect at the top-antitop threshold and translates into the decay products. Studying the charge asymmetries of the bottom-quarks and leptons from top decays gives therefore additional information on the chiral properties of new currents. For instance, an axigluon with purely right-handed couplings does not contribute to the charge asymmetry at tree level, because the radiated $W$ boson forces the initial-state quark current to be purely left-handed.

Due to the low cross section, the observability of the charge asymmetry in $t\bar t + W$ production relies on high statistics. With a luminosity of more than $300\,{\rm fb}^{-1}$ LHC data during run 2, a measurement of the SM rapidity asymmetry is expected to be feasible at the percent level. Contributions of new physics are generally larger than in $t\bar t$ and $t\bar t + j$, due to the dominance of quark-antiquark states in $pp\to t\bar t + W$. The charge asymmetry in $t\bar t + W$ production can thus be considered a complementary tool to distinguish between new phenomena in top-antitop production in the long-term run of the LHC.

\section{Acknowledgements}
I thank the organizers of FFP14 for inviting me to this conference and for travel support.

\end{document}